\documentclass[pra,superscriptaddress,floatfix,amsmath,amsfonts,twocolumn]{revtex4}  
\usepackage{enumitem}
\usepackage{amsmath}
\usepackage{graphicx}
\usepackage{graphics}
\usepackage{epsfig}
\usepackage{bm}
\usepackage{bbold}
\usepackage{amssymb}
\usepackage{color}

\newcommand {\tphi}{\tilde{\phi}}

\newcommand {\trho}{\tilde{\rho}}

\newcommand{\p}{{\cal P}}
\newcommand{\PT}{{\cal PT}}
\newcommand{\T}{{\cal T}}
\newcommand{\cO}{\mathcal{O}}
\newcommand{\cA}{{\cal A}}
\newcommand{\cB}{{\cal B}}
\newcommand{\cC}{{\cal C}}
\newcommand{\cL}{{\cal L}}

\newcommand{\cH}{{\cal H}}

\newcommand{\cK}{{\cal K}}

\newcommand{\tvarphi}{\tilde{\varphi}}

\newcommand{\bDelta}{ \mbox{\boldmath$\Delta$\unboldmath}}

\newcommand{\bZ}{{\bm Z}}

\newcommand{\px}{{\partial_x}}
\newcommand{\pxx}{{\partial_x^2}}
\newcommand{\pxxxx}{{\partial_x^4}}
\newcommand{\pz}{{\partial_z}}

\newcommand{\ra}{\rangle}
\newcommand{\la}{\langle}
\newcommand{\vphi}{\varphi}
\newcommand{\tvphi}{\tilde{\varphi}}
\newcommand{\sech}{{\operatorname{sech}}}
\newcommand{\RE}{{\operatorname{Re\,}}}
\newcommand{\IM}{{\operatorname{Im\,}}}

\begin{document}

\title{Metastable two-component solitons near an exceptional point}

\author{Dmitry A. Zezyulin}
\affiliation{ITMO University, St. Petersburg 197101, Russia}
\email{d.zezyulin@gmail.com}

\author{Yaroslav V. Kartashov}
\affiliation{Institute of Spectroscopy, Russian Academy of Sciences, Troitsk, Moscow, 108840, Russia}

\author{Vladimir V. Konotop}
\affiliation{Departamento de F\'{i}sica and Centro de F\'{i}sica Te\'orica e Computacional, Faculdade de Ci\^encias, Universidade de Lisboa, Edifício C8, Campo Grande,  Lisboa 1749-016, Portugal}
\email{vvkonotop@fc.ul.pt}

\date{\today}

\begin{abstract}
We consider a two-dimensional nonlinear waveguide with distributed gain and losses. The optical potential describing the system consists of an unperturbed complex potential depending only on one transverse coordinate, i.e., corresponding to a planar waveguide, and a small non-separable perturbation depending on both transverse coordinates. It is assumed that the spectrum of the unperturbed planar waveguide features an exceptional point (EP), while the perturbation drives  the system into the unbroken phase. Slightly below the EP, the  waveguide   sustains two-component envelope solitons. We derive one-dimensional equations for the slowly varying envelopes of the components and show their stable propagation. When both traverse directions are taken into account within the framework of the original model, the obtained two-component bright solitons become metastable and persist over remarkably long propagation distances.
\end{abstract}
\maketitle

\section{Introduction}

Stable propagation of linear waves in either conservative or dissipative systems requires reality of the spectrum of the governing evolution operator. When this operator is non-Hermitian and depends on control parameters, its spectrum can undergo qualitative changes upon variation of these parameters. In particular, the spectrum can change from purely real to a complex one. This (phase) transition between real and complex spectra typically occurs either through an exceptional point (EP) in the discrete spectrum~\cite{BenBoet1998,Bender,Heiss} or through a spectral singularity in the continuous spectrum~\cite{KZ2017,Yang2017}. Although EPs as well as spectral singularities are introduced as characteristics of linear spectral problems~\cite{Kato}, they also impact propagation of nonlinear waves. First of all, stability of linear waves of a given nonlinear system is a necessary (although not yet sufficient) condition for stability of localized nonlinear waves, for example, of bright solitons (see \cite{KYZ,Suchkov2016} for  review). An EP in the spectrum of the underlying linear system affects the equations governing weakly nonlinear waves having propagation constants in the vicinity of the EP~\cite{NiZY2012,NixYang}. If parameters of a nonlinear medium are close to an EP locally, i.e., only in a given spatial domain, a soliton interacting with such domain can be scattered according to different scenarios~\cite{BluHaHuKo2014}. In a waveguide geometry characterized by a separable optical potential (created by modulation of the dielectric permittivity) the existence of EP in the linear spectrum of the carrying transverse modes of a separable optical potential can change the sign of the effective Kerr nonlinearity felt by a wavepacket propagating along the waveguide~\cite{MidKon}.
 
In a more general context of nonlinear systems, an EP is sometimes introduced as a point of coalescence of the eigenvalues and eigenvectors of a {\em nonlinear} eigenvalue problem. Location of such EP in the parameter space depends on the nonlinearity, i.e., on the amplitude of the field. This was particularly well studied for models with double-well potentials~\cite{nonlinEP1,nonlinEP2,nonlinEP3}, and it was also found that the nonlinearity may have significant impact on the spectrum of the system in the vicinity of an EP. Thus, considering a nonlinear system with parameters, at which its linear limit is close enough to an EP, one can expect fragility of the stability that may be destructively affected by the nonlinearity. Therefore, one may expect considerable constraints on stable propagation of nonlinear waves in such systems. In this paper, we show that this is not necessarily so: metastable solitons can exist even when the parameters of the underlying linear system are in close proximity of an EP. 

The organization of our paper is as follows. In Sec.~\ref{sec:statement} we introduce  a nonlinear planar waveguide, which is described by an optical potential depending on one of the transverse directions and features   an EP. Then in Sec.~\ref{sec:perturbation}   we develop a perturbation theory for the spectrum of the corresponding non-Hermitian evolution operator near the EP in the presence of a nonseparable perturbation depending on both transverse coordinates. In Sec.~\ref{sec:equations} we derive the two-component system of equations governing evolution of the slowly varying amplitudes of the guided modes having the propagation constants in close proximity of the second-order EP. We employ the method of multiple-scale expansion where {\em two} coupled modes have to be accounted for self-consistency of theory. The obtained system for slowly varying amplitudes features high-order dispersion, resembling (although not coinciding with) dispersion that may lead to pyramid diffraction, considered previously in~\cite{NixYang}. In Sec.~\ref{sec:1Dsoliton} we describe solitons of the derived effective one-dimensional (1D) model. Such solitons are stable in the 1D model and they become metastable in the full 2D model governing light propagation in the dissipative waveguide (Sec.~\ref{sec:2D}).

\section{The model}
\label{sec:statement}

\begin{figure}
	\begin{center}
		\includegraphics[width=0.99\columnwidth]{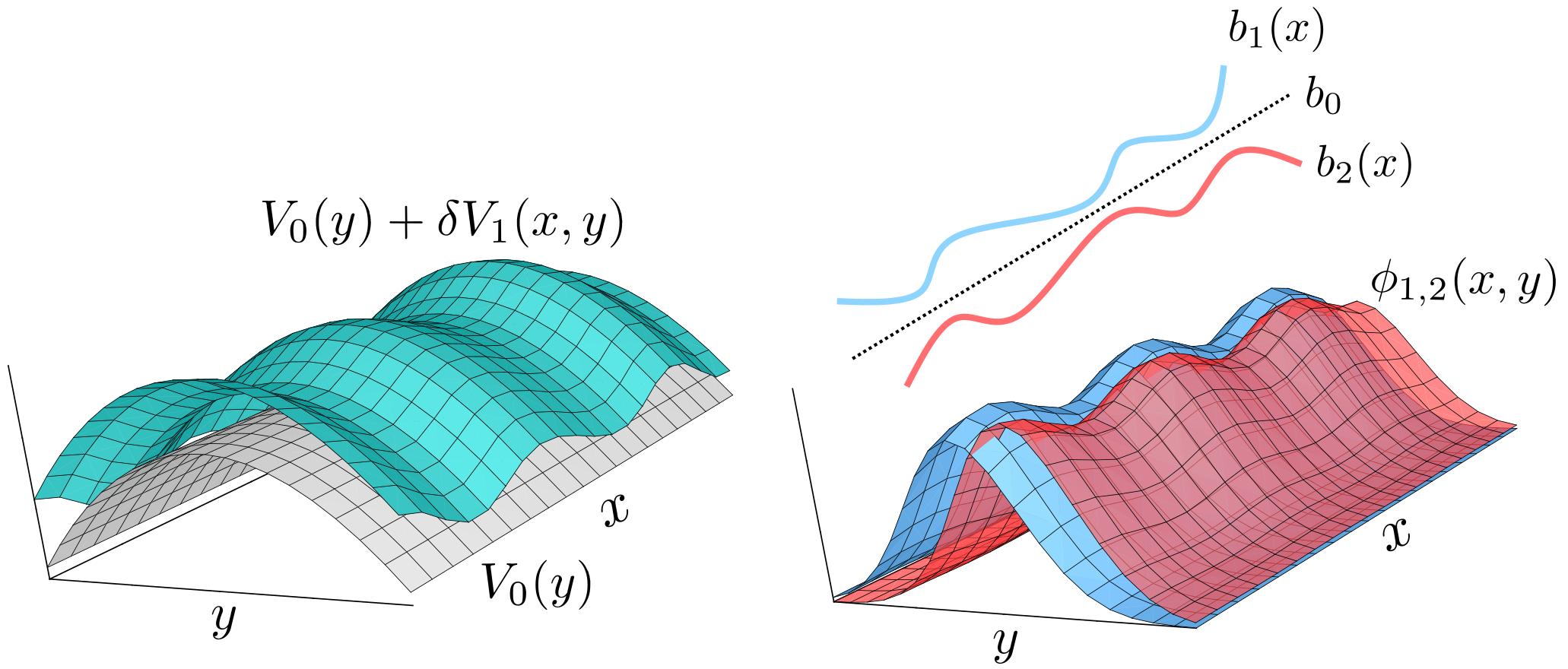}
	\end{center}
\caption{Schematics of the geometry. The  unperturbed potential $V_0(y)$ (left panel) features an EP at each value of $x$. For the perturbed potential $V_0(y)+\delta V_1(x,y)$ the degeneracy is lifted, and for each $x$ the waveguide operates slightly below the EP, which means that for each $x$ there are two close real-valued propagation constants $b_{1,2}(x)$ (right panel) and two eigenfunctions $\phi_{1,2}(x,y)$ which are about to merge but nevertheless are distinct. }
	\label{fig:idea}
\end{figure}

We consider propagation of a paraxial  beam along the $z$-direction in a medium with gain and losses modulated along $x$- and $y$-directions. The amplitude of the field $\Psi$ in dimensionless units is governed by the nonlinear Schr\"odinger equation  
\begin{equation}
\label{opt-eq1}
i\frac{\partial \Psi}{\partial z}= -\frac{1}{2}\nabla^2 \Psi-V_\delta(x,y)\Psi
+\chi |\Psi|^2\Psi,
\end{equation}
where $\nabla=(\partial_x,\partial_y)$, the complex-valued optical potential $V_\delta(x,y)$ is parameterized by a real control parameter $\delta$, and the real coefficient  $\chi$ characterizes Kerr nonlinearity of the medium. Considering $0\leq \delta\ll 1$, we assume that the potential  $V_\delta(x,y)$  can be represented in the form
\begin{eqnarray}
\label{eq:Vdelta}
	V_\delta (x,y)=V_0(y)+\delta V_1(x,y). 
\end{eqnarray}
Here $V_0(y)$ is a complex-valued potential for which the spectrum of the linear non-Hermitian Hamiltonian  
\begin{align}
\label{eq:H0}
	H_0:=-\frac{1}{2}\partial_y^2 - V_0(y)
\end{align}
has an EP $-b_0$ on the real axis, $b_0\in\mathbb{R}$, with $\phi_0(y)$ being the respective eigenfunction:
\begin{align}
	\label{phi0}
	H_0\phi_0=-b_0\phi_0.
\end{align} 
Schematics of the described waveguide is illustrated in Fig.~\ref{fig:idea}. Since  $x$ does not enter Eq.~(\ref{phi0}),  we can say that the waveguide has the EP at each value of $x$. 

For further consideration, we define the associated (generalized) eigenfunction $\vphi_0(y)$: 
\begin{align}
	\label{varphi0}
	(H_0+b_0)\varphi_0=\phi_0,
\end{align} 
as well as the eigenfunction   $\tphi_0$ and generalized eigenfunction $\tvarphi_0$ of the Hermitian conjugate $H_0^\dag$: 
 \begin{align}
	 H_0^\dagger\tphi_0=-b_0\tphi_0, 
	\qquad (H_0^\dagger+b_0)\tvarphi_0=\tphi_0.
\end{align}
Hereafter we use tildes for the spectral characteristics of the adjoint eigenvalue problem. In our case $H^\dagger=H^*$, and hence $\tphi_0=\phi_0^*$, $\tvphi_0=\vphi_{0}^*$, where asterisks mean    complex conjugation.  We emphasize that neither $\phi_0$ nor $\varphi_0$ depend on $x$; they depend only on $y$.

A signature of the EP is the orthogonality condition 
\begin{align}
\langle \tphi_0,\phi_0\rangle=0.
\end{align}
Hereafter we use the inner product defined as
\begin{align}
\langle f,g\rangle:=\int_{-\infty}^\infty f^*(x,y)g(x,y)dy.
\end{align}
Generally, the eigenfunction $\phi_0$  is defined up to an arbitrary nonzero coefficient, and the generalized eigenfunction  $\varphi_0$ is defined up to the addition of an arbitrary multiple of $\phi_0$. It will be convenient to fix two corresponding constants by imposing the following  normalization \cite{Malibaev2005}: 
\begin{eqnarray}
	\label{normal10}
	 \langle \tvarphi_0,\phi_0\rangle=\langle \tphi_0,\varphi_0\rangle=1,
	 \quad
\end{eqnarray}
and the orthogonality condition
\begin{eqnarray}
 \label{self-orthon}
	 \la\tvphi_0, \vphi_0\ra = 0.
\end{eqnarray}

\section{Perturbation theory for the spectrum near  EP}
\label{sec:perturbation}

Suppose, that $V_1(x,y)$ is chosen such that the spectrum of the perturbed linear Hamiltonian 
\begin{align}
	H_\delta:=H_0-\delta V_1(x,y)
\end{align}
 is all-real and there exist two eigenvalues $b_{1,2}(x)$, i.e., 
\begin{eqnarray}
\label{opt-eigen}
H_\delta \phi_{j}(x,y)=-b_{j}(x) \phi_j(x,y), \quad j=1,2,
\end{eqnarray}
satisfying the condition
\begin{eqnarray}
\label{eq:order}
	b_1(x)>b_0>b_2(x)
\end{eqnarray}
for all $x$. We recall that $x$ plays the role of a parameter in eigenvalue problem (\ref{opt-eigen}). We therefore can say that in the two-dimensional plane of parameters $(\delta,x)$ the Hamiltonian $H_\delta$   has an \emph{exceptional line} $\delta=0$. 

We also   define adjoint operator and eigenfunctions:
\begin{eqnarray}
	H_\delta^\dagger=-\frac{1}{2}\partial_y^2-[V_\delta (x,y)]^*=H_0^\dagger-\delta [V_1(x,y)]^*
\end{eqnarray}	
and
\begin{eqnarray}
\label{opt-eigen-conj}
	H_\delta^\dagger \tphi_{j}(x,y)=-b_{j}(x) \tphi_j(x,y), \quad j=1,2.
\end{eqnarray}
Furthermore, we assume that  $H_\delta$ does not have EPs when $\delta>0$ [this condition can be relaxed: what is really important for our consideration is the condition (\ref{eq:order})]. 
Then the following biorthogonality relations hold:  
\begin{align}
	\label{orthog}
	\langle \tphi_1,\phi_2\rangle=	\langle \tphi_2,\phi_1\rangle=0.
\end{align}

The structure of the eigenfunctions in the vicinity of the EP can be described in terms of the following asymptotic expansions   ($j=1,2$):
\begin{subequations}
	\label{eq:phi}
\begin{align}
	\label{phi1-0}
	 \phi_{j}=\phi_0+(-1)^j {\delta}^{1/2} \rho_j+\delta \rho_{j1}+{\delta^{3/2}\rho_{j2}} +\mathcal{O}(\delta^{2}),\\
\label{phi2-0}
	\tphi_{j}=\tphi_0+(-1)^j{\delta}^{1/2} \trho_j+\delta \trho_{j1}+{\delta^{3/2}\trho_{j2}}+ \mathcal{O}(\delta^{2}),
\end{align}
\end{subequations}
at $\delta\to 0$. In Eqs.~(\ref{eq:phi}), the functions  $\rho_j$ and $\trho_j$ are so-far undetermined corrections that generically depend on $x$ and $y$. Since in the eigenvalue problem (\ref{opt-eigen}) $x$ is a parameter, we can represent
\begin{align}
	 \label{b12} 
  	b_j(x)=b_0 -(-1)^j  {\delta}^{1/2} \beta_j(x) +\delta\beta_{j1}(x) 
 \nonumber 	\\ 
 + {\delta^{3/2}\beta_{j2}(x)} 	+  \mathcal{O}(\delta^{2}),
\end{align}
where the coefficients $\beta_j$, $\beta_{j1}$ and $\beta_{j2}$ are to be found. Substituting Eqs.~ (\ref{eq:phi})--(\ref{b12}) in (\ref{opt-eigen}), one observes that the obtained equation in the leading order $\delta^0$ is satisfied. The balance of $\delta^{1/2}$-order terms requires that ($j=1,2$)
\begin{align}
	(H_0+b_0)\rho_j=  \beta_j(x)\phi_0.  
\end{align}
Comparing this equation with  Eq.~(\ref{varphi0}) we conclude that
\begin{align}
	\label{c1c2}
	 \rho_{j}(x,y)= \beta_j(x)\varphi_0(y), \quad 
	  \trho_{j}(x,y)= \beta_j(x)\tvarphi_0(y).
\end{align} 
Further, from the orthogonality condition (\ref{orthog}) we obtain  the relation $ {\delta}^{1/2} (\beta_1-\beta_2)+\mathcal{O}(\delta)=0$.  Therefore  
\begin{align}
	\label{beta}
	\beta_1(x)=\beta_2(x)=:\beta(x),
\end{align}
 and hence
\begin{align}
	\label{chi12}
	\rho_1(x,y)=\rho_2(x,y)= \beta(x)\varphi_0(y).
\end{align}
Now Eqs.~(\ref{phi1-0}) and (\ref{phi2-0}) can be rewritten as
\begin{subequations}
\begin{align}
	\label{phi1-1}
	\phi_{j}=\phi_0+(-1)^j {\delta}^{1/2} \beta(x) \varphi_0+\mathcal{O}(\delta),
	\\
	\label{phi2-1}
		\tphi_{j}=\tphi_0+(-1)^j {\delta}^{1/2}\beta(x) \tvarphi_0+\mathcal{O}(\delta),
\end{align}
\end{subequations}
resulting in the normalization condition 
\begin{align}
	\label{normal_1}
	 \langle \tphi_{j},\phi_j\rangle=(-1)^j 2 {\delta}^{1/2} \beta (x) + \cO(\delta). 
\end{align}
To determine $\beta(x)$, we consider the next order, $\mathcal{O}(\delta)$, of equations (\ref{opt-eigen}):
\begin{align}
		(H_0+b_0)\rho_{j1} {-} V_1\phi_0 = \beta^2\varphi_0  { -}  \beta_{j1} \phi_0.
		\label{eq:delta12}
\end{align} 
Applying $\la \tphi_0, \cdot\ra$,  either for $j=1$ and $j=2$ we obtain
\begin{eqnarray}
\label{eq:beta^2}
	\beta^2(x)=-\langle \tphi_0, V_1{\phi}_0  \rangle.
\end{eqnarray}
Obviously, our analysis is meaningful only if the right-hand side of Eq. (\ref{eq:beta^2}) is positive. We also note that by choosing an appropriate $V_1(x,y)$, depending on both variables, $x$ and $y$, one can obtain any desirable function $\beta(x)$, If, however, $V_1$ does not depend on $y$, then $\beta(x)=0$. While Eq.~(\ref{eq:beta^2}) does not define the sign of $\beta$, the previously imposed convention (\ref{eq:order}) implies $\beta(x)>0$ (notice that the inequality is strict).

Under the condition (\ref{eq:beta^2}), a solution for  (\ref{eq:delta12}) reads
\begin{align}
\label{eq:chi1121}
\rho_{11} = f - \beta_{11}\varphi_0, \qquad \rho_{21} = f - \beta_{21}\varphi_0,
\end{align}
where $f$ solves the equation 
	\begin{align}
(H_0+b_0)f  =   V_1\phi_0  +  \beta^2\varphi_0.
\end{align} 
The function $f$ is defined up to the addition of an arbitrary multiple of $\phi_0$, but from the following analysis it will become evident that without loss of generality this multiple can be set to zero.

Proceeding to the  $\mathcal{O}(\delta^{3/2})$-order,  from equations (\ref{opt-eigen}) we obtain ($j=1,2$)
	\begin{align}
\label{eq:delta22}
(H_0+b_0)\rho_{j2}  = (-1)^j\beta (\rho_{j1} - \beta_{j1}\varphi_0  +  V_1 \vphi_{0}) -\beta_{j2} \phi_0.
\end{align} 
Solvability conditions for these equations read
\begin{align}
\beta_{j1} = \langle\tphi_0, V_1\varphi_0\rangle + \langle\tphi_0, \rho_{j1} \rangle.
\end{align}
Substituting here $\rho_{11}$ and $\rho_{21}$  from (\ref{eq:chi1121}), we obtain the next-order coefficients of the expansion for the propagation constants in the form:
\begin{align}
\label{eq:beta2-1}
\beta_{11} = \beta_{21} = \frac{1}{2} \langle\tphi_0, V_1\varphi_0\rangle + \frac{1}{2} \langle\tphi_0, f \rangle =: \gamma(x).
\end{align}
On the other hand, the orthogonality condition (\ref{orthog}) in the $\cO(\delta)$-order requires  
\begin{align}
\label{eq:beta2-2}
\gamma = \langle \tphi_0, f\rangle. 
\end{align}
Combining this expression with  (\ref{eq:beta2-1}), we obtain 
\begin{equation}
\label{eq:beta2}
\gamma = \la \tphi_0, V_1\vphi_0\ra.
\end{equation}
The requirement for the propagation constant to be real, implies that the perturbation $V_1$ should be chosen to ensure the reality of the right-hand side of (\ref{eq:beta2}). If this condition is satisfied, then (\ref{eq:beta2})  and the relations
\begin{align}
	\label{eq:orthochi}
		\langle \rho_{j1}, \tphi_0\rangle  = \la V_1\phi_0, \tvarphi_0\ra  		 - \beta_{j1},
\end{align} 
obtained by applying $\langle \cdot, \tvarphi_0\rangle$ to (\ref{eq:delta12}), yield
\begin{align}
\langle \rho_{j1}, \tphi_0\rangle =  
\langle \px \rho_{j1}, \tphi_0\rangle  = \langle \pxx \rho_{j1}, \tphi_0\rangle  = 0.
\end{align}
Therefore, the estimate~(\ref{normal_1}) can be improved as follows
\begin{align}
\langle \tphi_{j},\phi_j\rangle=(-1)^j 2 {\delta}^{1/2} \beta (x) + \cO(\delta^{3/2}). 
\end{align}
Finally, with the same accuracy we compute useful relations
\begin{align}
\la \tphi_k, \px \phi_j\ra  =  (-1)^j \beta_x \delta^{1/2} + \cO(\delta^{3/2}),\\[1mm]
\la \tphi_k, \pxx \phi_j\ra  =  (-1)^j \beta_{xx} \delta^{1/2} + \cO(\delta^{3/2}).
\end{align}

\section{Multiple-scale expansion}
\label{sec:equations}

Now we turn to the nonlinear model and, using the multiple-scale expansion, look for the solution of Eq.~(\ref{opt-eq1}) in the form
\begin{align}
\label{eq:Psi}
	\Psi=\delta^{1/2}e^{ib_0z}
	\left[
	\phi_{1}(x,y)U_1(x,z) 
	+\phi_{2}(x,y)U_2(x,z)
 	\right]
 		\nonumber \\ 
 		+\mathcal{O}(\delta^{3/2}),	
\end{align}
where $U_{1,2}(x,z)$ are the envelopes of the two modes that coalesce in the EP in the limit $\delta=0$. Thus, the field we are looking for is two-component. We substitute (\ref{eq:Psi}) into the main equation (\ref{opt-eq1}) and apply $\la \tphi_1, \cdot\ra$ and $\la \tphi_2, \cdot\ra$ to the resulting expression. Using the results of Sec.~\ref{sec:perturbation}, we arrive at a system of two coupled equations that govern the dynamics of $U_1$ and $U_2$:
\begin{align}
	\label{opt-total-5a}
	i\partial_z U_1=&\cH_0 U_1 {-} \delta^{1/2}\beta U_1   {+\frac{\beta_{xx}}{4\beta}U_2}
	- \frac{\beta_x}{2\beta}  \partial_x U_1 + \frac{\beta_x}{2\beta}  \partial_x U_2 
	\nonumber \\ 
	&- \delta^{1/2}\frac{\chi_0}{2\beta} 
	\left(U_1+U_2\right)^2\left(U_1^*+U_2^*\right), 
	\\ 
	\label{opt-total-5b}
	i\partial_z U_2=&\cH_0U_2 {+}\delta^{1/2}\beta U_2  {+\frac{\beta_{xx}}{4\beta} U_1}
	+  \frac{\beta_x}{2\beta} \partial_x U_1  -  \frac{\beta_x}{2\beta}  \partial_x U_2 
	\nonumber \\ 
&	+\delta^{1/2}\frac{\chi_0}{2\beta} \left(U_1+U_2\right)^2\left(U_1^*+U_2^*\right). 
\end{align}
Here
\begin{align}
		\label{eq:cH0}
	\cH_0=-\frac{1}{2}\partial_x^2-\frac{\beta_{xx}}{{4}\beta}, 
\end{align}
the effective nonlinearity is determined as
\begin{align}
	\label{chi0}
	\chi_0=\chi \langle  \phi_0^*,|\phi_0|^2\phi_0\rangle, 
\end{align}
and all terms of the order of $\delta$ (and higher) are neglected. Upon derivation of the system (\ref{opt-total-5a})--(\ref{opt-total-5b}) we used that  ($i,j,k=1,2$)
\begin{align*}
	\langle \tphi_i,|\phi_j|^2\phi_k\rangle  
	=\langle  \phi_0^*,|\phi_0|^2\phi_0\rangle +\mathcal{O}({\delta}^{1/2}),
	\\
	\langle \tphi_i,\phi_j^2\phi_k^*\rangle  
	=\langle  \phi_0^*,|\phi_0|^2\phi_0\rangle  +\mathcal{O}({\delta}^{1/2}).
\end{align*}

Generally speaking, the effective nonlinearity coefficient $\chi_0$ obtained in  (\ref{chi0}) is complex-valued. However, it is necessarily real if the unperturbed potential is $\p_y\T$ symmetric. Here,  according to standard definitions, the operator $\p_y$ corresponds to the reversal of the $y$-axis, and the operator $\T$ corresponds to the complex conjugation, and thus $[\p_y\T, H_0]=0$, where the unperturbed operator $H_0$ is defined in (\ref{eq:H0}). Indeed, the $\p_y\T$ symmetry implies that  
$\p_y\T \phi_0= e^{i\theta}\phi_0$, where $\theta$ is a constant phase. From (\ref{varphi0}) and  (\ref{normal10}) it follows that $e^{2i\theta}=1$, i.e., possible values of $\theta$ are $0$ and $\pi$. In either case  $\p_y|\phi_0|^2=|\phi_0|^2$, and 
one can  verify that $\chi_0$ is real:
\begin{align*}
	\langle  \phi_0^*,|\phi_0|^2\phi_0\rangle 
	&= \T \langle  \phi_0,|\phi_0|^2\T\phi_0\rangle 
	\nonumber  \\
	&=  \langle e^{-i\theta}\p_y\T  \phi_0,|\phi_0|^2e^{i\theta}\p_y\phi_0\rangle^* 
	\nonumber \\
	&= e^{-2i\theta}\langle \T  \phi_0,|\phi_0|^2\phi_0\rangle^*
	= \langle  \phi_0^*,|\phi_0|^2\phi_0\rangle^*.
\end{align*}

Importantly, the effective nonlinearities for different components of the field, described by (\ref{opt-total-5a})--(\ref{opt-total-5b}), have opposite signs. This effect resembles the finding reported in \cite{MidKon} where it has been shown that the presence of an EP in the spectrum of the underlying linear problem can change the sign of the effective nonlinearity.

Equations (\ref{opt-total-5a})--(\ref{opt-total-5b}) acquire a more convenient form if one introduces new functions $	U_{\pm}=U_2\pm U_1$ which satisfy the following system:
\begin{subequations}
	\label{opt-total-7}
\begin{align}
\label{opt-total-7a}
i\partial_z U_+=& -\frac{1}{2}\pxx U_+ + \delta^{1/2}\beta U_-,
\\ 
\label{opt-total-7b}
i\partial_z U_-=&\left(-\frac{1}{2}\pxx -  \frac{\beta_{xx}}{2\beta}\right)U_-  
-  \frac{\beta_x}{\beta}  \partial_x U_-   
\nonumber \\ 
& +\delta^{1/2}\beta U_+ 
+\delta^{1/2}\frac{\chi_0}{\beta} |U_{+}|^2U_+.
\end{align}
\end{subequations}

The equations obtained for the envelopes make explicit the scaling of the solutions as well as constraints that should be imposed on the $\beta(x)$ dependence. Indeed, from Eq.~(\ref{opt-total-7a}) we conclude that the envelope is smooth, in the sense that it depends on the scaled variables $\delta^{1/2}z$ and $\delta^{1/4}x$.  From Eq.~(\ref{opt-total-7b}) it follows that the consistency of the multiple-scale expansion requires $|\beta_x/\beta|\lesssim \delta^{1/4}$ and $|\beta_{xx}/\beta|\lesssim \delta^{1/2}$.

\section{1D solitons  	at constant $\beta$}
\label{sec:1Dsoliton}

Let us now consider solitons of the 1D model (\ref{opt-total-7}) at a constant $\beta>0$, when $\beta_x=\beta_{xx}=0$.
For stationary solutions, $U_\pm  = u_\pm(x)  e^{i\mu z}$, the system (\ref{opt-total-7}) reduces  to 
\begin{subequations}
		\label{opt-total-stat}
\begin{align}
	\label{opt-total-stat1}
	\mu u_+=\frac{1}{2}\frac{d^2 u_+}{dx^2}&- \delta^{1/2}\beta u_-, 
	\\ 
	\label{opt-total-stat2}
	\mu u_-=\frac{1}{2}\frac{d^2 u_-}{dx^2} &-\delta^{1/2}\beta u_+ 
 	-\delta^{1/2}\frac{\chi_0}{\beta} |u_{+}|^2u_+.
\end{align} 
\end{subequations}
The spectrum  of the linear ($\chi_0=0$) limit of this system ($U_\pm\propto e^{i\mu z+i k x}$) has two branches
\begin{align}
	\label{eq:continuum}
	\mu_\pm=-\frac{k^2}{2}\pm\sqrt{\delta}\beta. 
\end{align}
 
System (\ref{opt-total-stat})  can be further reduced to a fourth-order  nonlinear equation 
\begin{align}
	\label{4equation}
	-\frac{1}{4}\frac{d^4u_+}{dx^4}+\mu \frac{d^2 u_+}{dx^2}+(\delta\beta^2-\mu^2)u_+ + \delta\chi_0|u_+|^2u_+=0.
\end{align}
Two comments are in order. First, one can see that in the vicinity of the EP governing equation includes the fourth-order dispersion, that corroborates the previous results on linear diffraction~\cite{NiZY2012,NixYang}. Second, the models similar to model (\ref{4equation}) can be encountered in fiber optics in description of evolution of pulses close to the zero-dispersion wavelength (see e.g.~\cite{Akhmediev,Hook}).

\subsection{Two-component solitons}

Equation (\ref{4equation}), and hence system (\ref{opt-total-stat}), allow for an exact solution. Indeed, for $\chi_0>0$ and  the propagation constant  
\begin{eqnarray}
\label{eq:mu1}
	 \mu=\frac{5}{3}\delta^{1/2}\beta>\mu_+
\end{eqnarray}
belonging to the semi-infinite gap of linear spectrum, one obtains~\cite{Akhmediev}
\begin{subequations}
	\label{eq:sol1}
\begin{eqnarray}
u_{+}(x) &=&  \beta\sqrt{\frac{10}{3\chi_0}}\sech^2\xi,\\
u_{-}(x)&=& -\beta\sqrt{\frac{10}{3\chi_0}} (\sech^2\xi+\sech^4\xi),
\end{eqnarray}
\end{subequations}
where for compactness we have introduced
\begin{align}
\xi = \sqrt{\frac{\beta}{3}} \delta^{1/4}x. \label{eq:xi}
\end{align}

\begin{figure}
	\begin{center}
		\includegraphics[width=0.99\columnwidth]{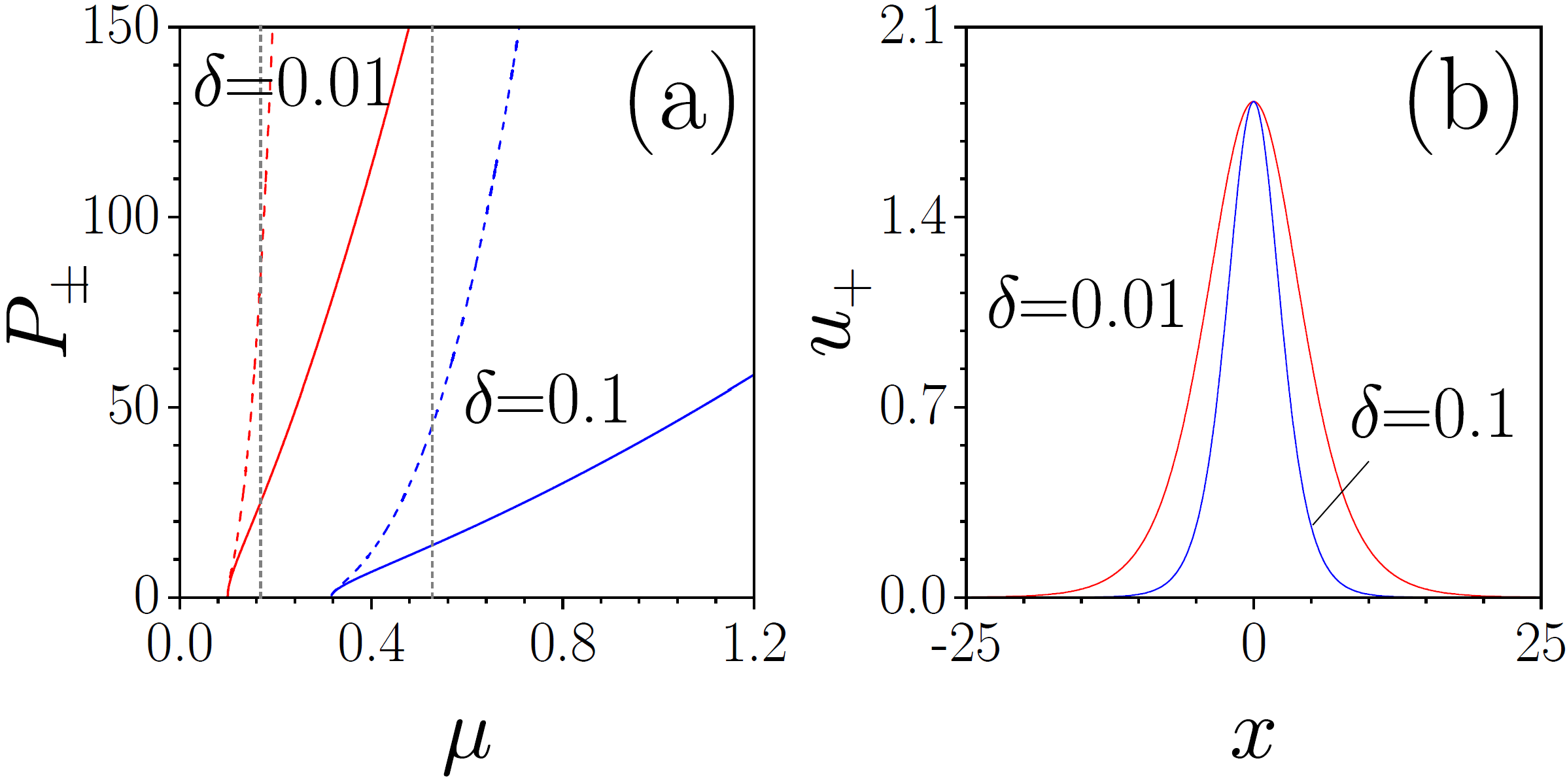}
	\end{center}
	\caption{(a) Families of solitons,  in terms of the dependencies of powers $P_{\pm}$ on $\mu$,  for $\delta=0.01$ (red curves) and $\delta=0.1$ (blue curves). Solid and dashed lines correspond to $P_+$ and $P_-$, respectively. Gray vertical lines denote values of $\mu$ that correspond to the exact solution (\ref{eq:mu1})--(\ref{eq:sol1}). (b) Profiles of corresponding exact solutions at $\mu=0.167, \delta=0.01$ (red curve) and $\mu=0.527, \delta=0.1$ (blue curve). Here $\chi_0=1$, $\beta=1$.}
	\label{fig:01}
\end{figure}

The exact soliton given by Eq.~(\ref{eq:sol1}) belongs to a continuous family parameterized by the propagation constant $\mu$. Examples of such families are shown in Fig.~\ref{fig:01}(a)  in the form of dependencies $P_\pm(\mu)$, where $P_\pm$ are   the powers carried by each of the  two components:
\begin{align}
	\label{P}
P_\pm = \int_{-\infty}^\infty |u_\pm(x)|^2 dx.
\end{align}

\subsection{Linear stability of two-component solitons}
Now  we proceed to the linear-stability analysis of the found solitons in the framework of the  two-component  model (\ref{opt-total-7}). For perturbed solutions in the form $U_\pm = e^{i\mu z}[u_\pm(x) + q_\pm(x,z)]$, where $q_\pm$ are small perturbations of real-valued solutions $u_\pm(x)$, the linearization of the two-component system (\ref{opt-total-7}) with constant $\beta$ gives the following eigenvalue problem after splitting into real and imaginary parts: 
 \begin{equation}
  \begin{array}{rcl}
 \pz \RE q_+ &=& \cL_0 \IM q_+ + \delta^{1/2}\beta\IM q_-,\\[2mm]
 \pz \IM q_+ &=& -\cL_0\RE q_+  -  \delta^{1/2}\beta\RE q_-,\\[2mm]
 \pz \RE q_- &=& \cL_0\IM q_- +\delta^{1/2}(\beta+ {\chi_0}{\beta^{-1}} u_+ ^2 )\IM q_+,\\[2mm]
 \pz \IM q_- &=& -\cL_0\RE q_- - \delta^{1/2}(\beta+ {3\chi_0}{\beta^{-1}} u_+ ^2 )\RE q_+,
 \end{array}
 \end{equation}
 where $\cL_0 = -(1/2) \pxx + \mu$.
 Making the substitution $(\RE q_+, \IM q_+, \RE q_-, \IM q_-) =   e^{\omega z}(Q_+, R_+, Q_-, R_-)$, where $\omega$ is the linear stability  eigenvalue (positive real part of $\omega$ corresponds to the exponential growth of the perturbation along the propagation distance), we then 
 eliminate $Q_-$ and $R_-$ from the linear stability equations. This leads to a \emph{quadratic} eigenvalue problem 
 \begin{align}
 	\label{eq:quadr}
 	(\omega^2\mathbb{1} + \omega \cC + \cK)\bDelta=0,
 \end{align} 
where
\begin{align}
	\bDelta = (Q_+, R_+)^T,
	\\[1mm]
  	\cC = \left(
 	\begin{array}{cc}
 		 {0}&-2\mu+\pxx  \\
 		2\mu-\pxx & 0
 	\end{array}
 	\right), \quad \cK = \left(
 	\begin{array}{cc}
 		\cL_+ &  {0}\\
 		{0} &  \cL_-
 	\end{array}
 	\right),\\[1mm]
 	\cL_\pm = -\frac{1}{4}\pxxxx + \mu \pxx + (\beta^2\delta - \mu^2) + (2\pm 1)\delta \chi_0 u_+^2,
  \end{align}
and $\mathbb{1}$ is the identity operator.  Quadratic eigenvalue problem (\ref{eq:quadr}) can be  further converted  into the generalized eigenvalue problem   \cite{quadratic}
 \begin{align}
 \label{eq:general}
(\cA -\omega \cB)\bZ =0,
 \end{align}
 where the augmented matrices   read
 \begin{align*}
 	\cA = \left(
 	\begin{array}{cc}
 		\cK & 0 \\
 		0 & -\mathbb{1}
 	\end{array}\right), \quad 
 	\cB = \left(
 	\begin{array}{cc}
 		-\cC & -\mathbb{1} \\
 		-\mathbb{1} & 0
 	\end{array}\right), \quad 	\bZ = \left(
 	\begin{array}{c}
 	\bDelta \\
\omega\bDelta
 	\end{array}\right).
 \end{align*}
 
 Numerical solution of the generalized eigenvalue problem (\ref{eq:general})   indicates that the soliton families with $\chi_0>0$ are entirely stable.  

\subsection{Embedded soliton}

For $\chi_0<0$, Eq.~(\ref{4equation}) admits  another exact bright soliton solution which can be written down as   \cite{Hook}
\begin{subequations}
\label{eq:sol2}
\begin{eqnarray}
u_+(x) &=&   {\beta}\sqrt{\frac{10}{3|\chi_0|}}\sech\xi\tanh\xi,\\
u_-(x) &=&    \beta \sqrt{\frac{10}{3|\chi_0|}}\sech\xi\tanh^3\xi,
\end{eqnarray}
\end{subequations}
where $\xi$ is given by (\ref{eq:xi}).
The propagation  constant of this solution
\begin{align}
	\mu=-\frac{5}{6}\sqrt{\delta\beta^2} 
\end{align}
belongs to the continuous spectrum, i.e., this is an embedded soliton which can hardly be expected to be stable.  The   instability of this solution has   indeed been confirmed by the linear-stability analysis (described in the previous section), as well as by using direct numerical simulations of one-dimensional propagation governed by vector model (\ref{opt-total-7}). Nevertheless, the fact that the system derived here simultaneously supports bright solitons in both focusing and defocusing media is rather interesting.

\section{Metastable 2D solitons}
\label{sec:2D}

Now we turn to the two-dimensional   solitons supported by the original model (\ref{opt-eq1}).  To find stationary solutions, we use the substitution $\Psi(x,y,z) = e^{i\lambda z} w(x,y)$. Comparing this substitution with (\ref{eq:Psi}), we obtain the approximate relation $\lambda \approx b_0 + \mu$, which connects the propagation constant of the 2D solitons (i.e., $\lambda$) with that of the 1D solitons considered above in Sec.~\ref{sec:1Dsoliton}. Using exact solutions obtained above for  the 1D model, one can produce reasonable analytical approximation for the 2D soliton profile. At the same time, feasibility of the experimental observation of such 2D solutions depends on the existence of optical potentials $V_\delta(x,y)$ featuring an EP at $\delta=0$ and purely real spectrum at $0<\delta\ll 1$. Examples of such potentials are well-known. We discuss two possible examples in the following subsections.

\subsection{Exactly solvable $\PT$-symmetric Scarff~II potential}

Let $V_\delta(x,y)$ be the $\p_y\T$-Scarff potential  \cite{scarff1,scarff2,scarff3}. For the analysis of its EP we use the representation (\ref{eq:Vdelta}) with
\begin{align}
\label{eq:scarffV0}
V_0 &=  \left[\frac{1}{4}(4\beta + 1)^2 - \frac{1}{8}\right]   \sech^2 y +  \frac{i}{4}(4\beta+1)^2 \sech y \tanh y,\\
V_1 &=  \frac{\sech^2 y}{ 2}.
\end{align}
For any  positive   $\beta>0$, the potential $V_0(y)$ defined by (\ref{eq:scarffV0}) is exactly at the EP that corresponds to the coalescence of two eigenmodes at the propagation constant $b_0= 2\beta^2$. The perturbation $\delta V_1$ increases the real part of total potential $V_\delta$ and therefore drives the system below the phase transition threshold.

Thanks to the solvability of the $\PT$-symmetric Scarf~II potential~\cite{scarff1,scarff2,scarff3}, exact expressions for the propagation constants are available   
\begin{align}
b_{1,2} = 
\frac{1}{8}\left[\sqrt{(4\beta+1)^2  + \delta} \pm \sqrt{\delta} - 1\right]^2\nonumber \\  =2\beta^2 \pm \beta \sqrt{\delta} + \gamma \delta   + \cO(\delta^{3/2}), \label{eq:b12scarff}
\end{align}
where [see Eqs.~(\ref{eq:beta2-1})--(\ref{eq:beta2})]
\begin{equation}
\label{eq:gammascarff}
\gamma = \frac{\beta}{2(4\beta+1)} + \frac{1}{8}.
\end{equation}
The eigenfunctions at the EP  read 
\begin{align}
\label{eq:phi0scarff1} 	
\displaystyle
\phi_{0} &= \frac{2^{2\beta+1}i\beta}{\sqrt{\pi}}  \exp\left\{ i\frac{4\beta+1}{2} \arctan\sinh y \right\}\sech^{2\beta }y,\\[2mm]
\label{eq:phi0scarff2}
\displaystyle
\varphi_0 &= \frac{\phi_0}{2\beta}(i\arctan\sinh y - \ln\sech y - \psi(1)+\psi(4\beta) - \ln 2),  
\end{align}
where $\psi(z) = d(\ln\Gamma(z))/dz = \Gamma'(z)/\Gamma(z)$ is the digamma function \cite{Olver}. One can check that the normalization conditions (\ref{normal10}) are satisfied. 
The nonlinear coefficient defined by (\ref{chi0}) is computed as
\begin{align}
\chi_0 = -\chi \frac{16\beta^3}{3\pi\mathrm{B}(6\beta,2\beta)},
\end{align}
where $\mathrm{B}(\cdot,\cdot)$ is the beta function. Thus $\chi_0>0$ ($\chi_0<0$) corresponds to the focusing (defocusing) nonlinearity of the physical model (\ref{opt-eq1}).

\begin{figure}
	\begin{center}		
		\includegraphics[width=0.99\columnwidth]{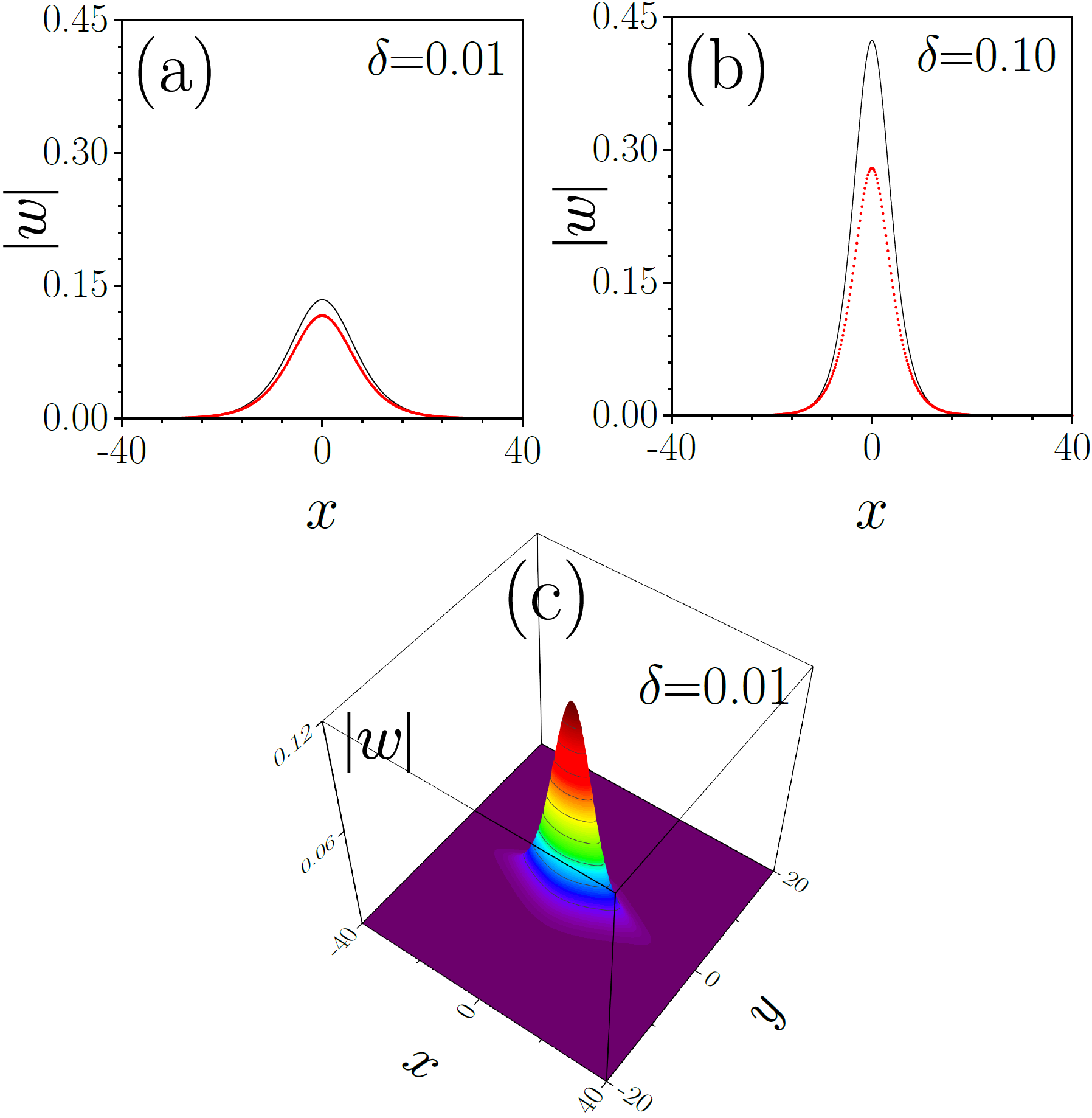}
	\end{center}
	\caption{(a,b) Comparison between analytical (black solid line) and numerical (red dots) soliton profiles in the $\PT$-Scarff potential for two values of $\delta$. Here we show amplitudes of the solutions at fixed $y=0$ as   functions of $x$. (c) Three-dimensional plot the of the numerical solution amplitude $|w(x,y)|$. In all   panels $\beta=0.4$, $\chi=-1$.}
	\label{fig:scarff}
\end{figure}

In Fig.~\ref{fig:scarff}(a,b) we compare analytical prediction for soliton shape (its cross-section at $y=0$) obtained using the combination of the exact 1D solution (\ref{eq:sol1}) and eigenfunctions given by (\ref{eq:phi0scarff1})--(\ref{eq:phi0scarff2}) with numerically obtained 2D soliton of Eq.~(\ref{opt-eq1}) having the same propagation constant $\lambda = 2\beta^2 + 5\delta^{1/2}\beta/3$. One can see that analytical and numerical solutions are very close at sufficiently small $\delta$ values, while with increase of $\delta$ the difference between them gradually increases.

\subsection{Numerical results in the  parabolic $\PT$-symmetric  potential}

For a more systematic study of the families of the 2D solitons, we choose a less sophisticated potential in the form  
\begin{align}
\label{eq:y^2}
V_0  = -y^2 +  i \gamma_0  ye^{-y^2/2}.  
\end{align}
One-dimensional nonlinear modes in a potential of similar form have been considered in \cite{Achi2012}. This potential has an EP at $\gamma_0\approx 2.1684$. Analytical expression for the eigenfunctions $\phi_0$ and $\vphi_{0}$ are not available in this case, but they can be found numerically.   In order to drive potential $V_0$  to the unbroken $\PT$-symmetric phase, we perturb it by decreasing the gain-and-loss amplitude by means of the following perturbation 
\begin{align}
V_1 =  -i\frac{\beta^2}{0.5521} ye^{-y^2/2}.
\end{align}
Notice that the numerical coefficient  in the denominator is chosen to ensure (\ref{eq:beta^2}). 

\begin{figure}
	\begin{center}		
		\includegraphics[width=0.99\columnwidth]{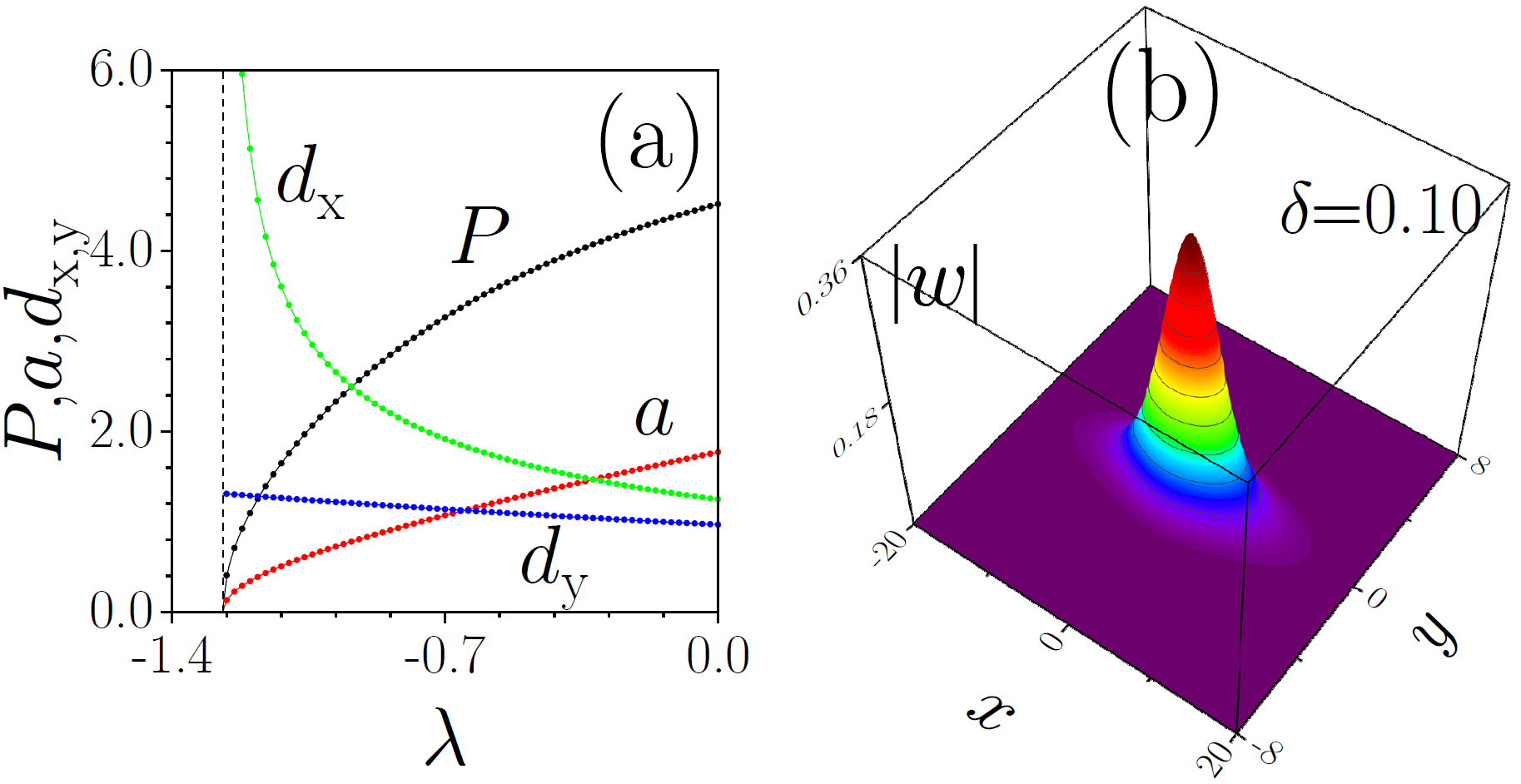}%
	\end{center}
	\caption{(a) Dependencies of power $P$, peak amplitude $a$, and widths $d_x$, $d_y$ on 		propagation constant $\lambda$ of 2D solitons  in parabolic $\PT$-symmetric potential at $\beta =1$ and  		$\delta =0.1$. (b) Profile of stable soliton at $\lambda =-1.2$. The focusing nonlinearity coefficient $\chi=-1$.}
	\label{fig:family}
\end{figure}

A family of 2D solitons obtained numerically  is shown in Fig.~\ref{fig:family}, where we present our results for the total power of solitons [cf.~(\ref{P})] 
\begin{align}
P=\iint |w(x,y)|^2dxdy,	 
\end{align} 
and for the soliton amplitudes and widths along the $x$ and $y$ axes. All these characteristics are functions of the propagation constant $\lambda$. While the analytical expression (\ref{eq:Psi}) is valid for soliton amplitudes $a\sim \delta^{1/2}$, the numerical continuation allows to obtain even large-amplitude solitons. The cutoff value [shown with dashed vertical line in Fig.~\ref{fig:family}(a)] corresponds to the edge of the continuum of two-dimensional scattering states. At the cutoff propagation constant the soliton amplitude and power vanish. At the same time, the widths $d_x$, $d_y$ illustrate the anisotropic nature of 2D solitons in our waveguide: as $\lambda$ approaches the cutoff, the soliton width in the $x$-direction diverges, while the width in the $y$-direction remains finite.

\begin{figure}
 	\begin{center}		
 		\includegraphics[width=0.99\columnwidth]{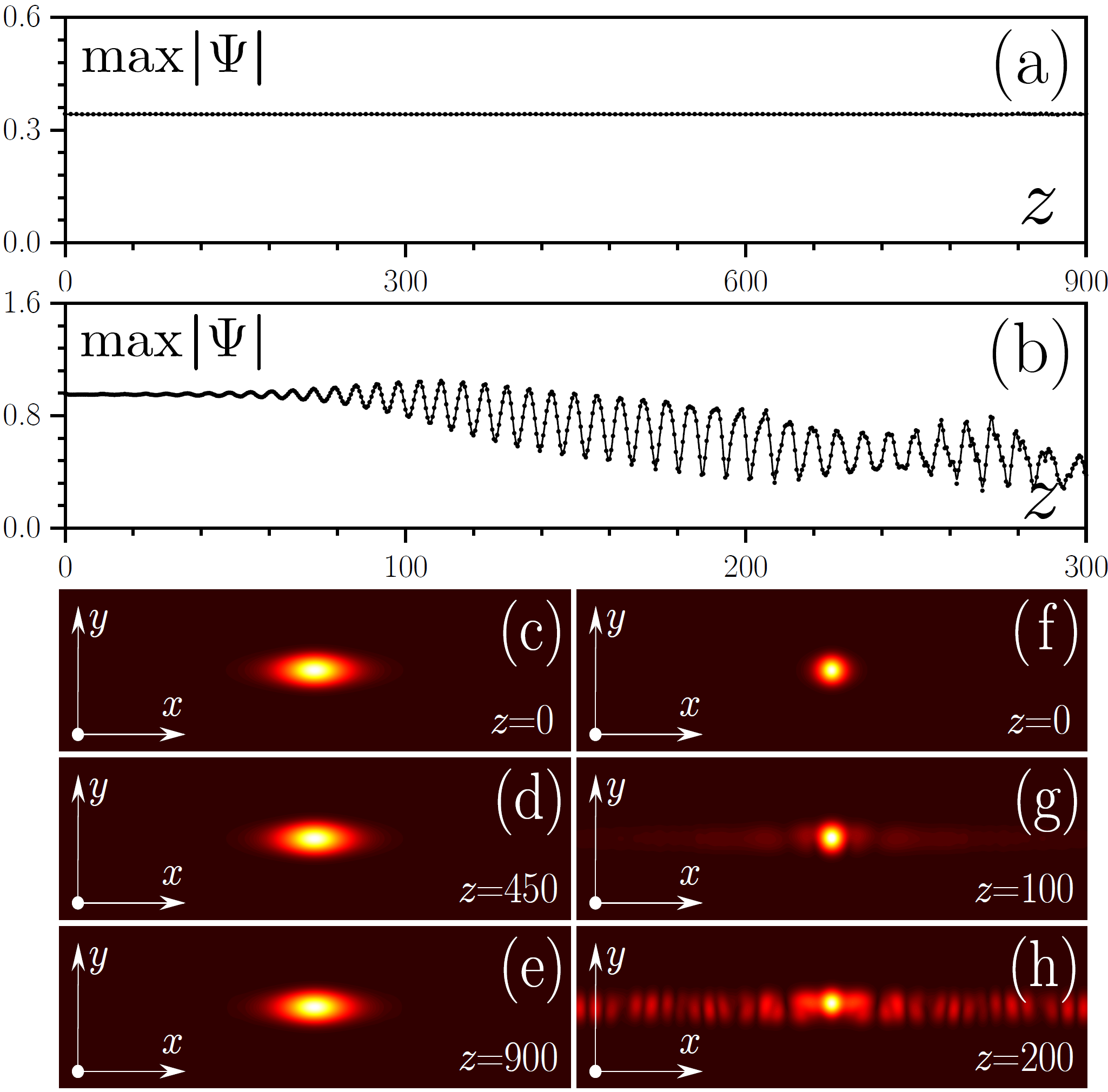}%
 	\end{center}
 	\caption{Peak amplitude versus distance and field modulus distributions at different propagation distances for stable 2D soliton with $\lambda =-1.2$ (a) and (c)--(e), and unstable  2D		soliton with $\lambda =-0.8$ (b) and (f)--(h) in parabolic $\PT$ potential. In all cases $\beta=1$ and  	$\delta=0.1$. Panels (c-h) correspond to spatial window $(x,y)\in [-40, 40]\times [-8, 8]$.  The focusing nonlinearity coefficient $\chi=-1$.}
 	\label{fig:meta}
 \end{figure}

While stability analysis performed in Sec.~\ref{sec:1Dsoliton} in the frame of the reduced 1D model for slowly varying envelopes has indicated that the 1D solitons are stable, this result does not yet guarantee stable  propagation of the respective 2D solitons constructed using (\ref{eq:Psi}). A systematic numerical study of soliton propagation governed by the $(2+1)$D equation (\ref{opt-eq1}) indicates that near the cutoff value the 2D solitons are robust and propagate over considerable distances without noticeable distortions even in the presence of input perturbations, but far from the cutoff the oscillatory instabilities come into play, whose strength gradually increases with the increase of soliton amplitude and propagation constant. The example of metastable evolution of the 2D soliton is presented in Fig. \ref{fig:meta}(a) showing that amplitude of such state remains practically unchanged with distance $z$, while cross-sections at different distances are shown in Fig. \ref{fig:meta}(c)-(e). The example of instability development for high-amplitude soliton stimulated by small input noise is presented in Fig. \ref{fig:meta}(b) and (f)-(h). As one can see, such unstable soliton starts radiating and at sufficiently large distance this radiation grows in amplitude and extends practically over the entire $x$ cross-section. We notice that weak oscillatory instabilities that affect propagation of the 2D solitons in our system can be possibly attributed to poorly localized (in the $x$ direction) unstable modes that bifurcate from the interior of the two-dimensional continuum \cite{Borisov}. However, an accurate analysis of this issue requires a separate and more detailed study.

\section{Conclusion}

In this work, we have shown that in a waveguide with gain and loss it is possible to obtain  propagation of metastable two-dimensional solitons with propagation constants in the vicinity of the exceptional point (but belonging to the unbroken phase). Such solitons are effectively two-component. Analytically they are described by the coupled linear and nonlinear Schr\"odinger equations which govern envelopes of the two carrier modes. The deviation of the propagation constant from the exceptional point is the small parameter of the multiple-scale expansion. The effective one-dimensional equations for the envelope allow for  exact bright soliton solutions for either sign of the nonlinearity coefficient. The envelope solitons are   stable in the one-dimensional setting, although they become metastable in the fully two-dimensional model.  The lifetime of such solitons is very large making them feasible for the experimental observation.

 \begin{acknowledgments}
The work of DAZ was supported by the Foundation for the Advancement of Theoretical Physics and Mathematics ``BASIS'' (Grant No. 19-1-3-41-1). 
VVK acknowledges financial support from the Portuguese Foundation for Science and Technology (FCT) under Contract no. UIDB/00618/2020.
 \end{acknowledgments}

\end{document}